\documentclass[11pt]{article}
\usepackage{amssymb}
\usepackage{amsfonts}
\usepackage{graphicx}
\usepackage{amsmath}

\makeatletter \@addtoreset{equation}{section} \makeatother

\newtheorem{theorem}{Theorem}
\newtheorem{lemma}{Lemma}

\newtheorem{remark}{Remark}

\newtheorem{proposition}{Proposition}

\begin{document}

\title{Fluctuations of eigenvalues of matrix models and their applications}

\author{T. Kriecherbauer$^\dagger$, M. Shcherbina$^\ddagger$
\\
\\
$^\dagger$Ruhr University of Bochum\\
\\
$^\ddagger$Institute for Low Temperature Physics Ukr.Ac.Sci
}

\date{}

\maketitle

\begin{abstract}
We study the expectation of  linear eigenvalue statistics of matrix models with any
$\beta>0$,
assuming that the potential $V$ is a real analytic function and that the corresponding equilibrium
measure has a one-interval support. We obtain the first order (with respect to $n^{-1}$) correction terms  for
the expectation and apply this result to prove bulk universality for real symmetric and symplectic
 matrix models with the same $V$.
\end{abstract}

\section{Introduction and main results}\label{s:1}

We consider ensembles of $n\times n$  real symmetric, hermitian or symplectic matrices $M$
 with the probability distribution
\begin{equation}\label{p(M)}
P_{n}(M)dM=Z_{n,\beta}^{-1}\exp \{-\frac{n\beta}{2}\mathrm{Tr}V(M)\}dM,
\end{equation}
where $\beta=1,2,4$ corresponds  to real symmetric, Hermitian, and symplectic case
respectively, $Z_{n,\beta}$ is a normalization  constant, $V:\mathbb{R}\to \mathbb{R%
}_{+}$ is a H\"{o}lder function satisfying the condition
\begin{equation}\label{condV}
V(\lambda )\ge 2(1+\epsilon )\log(1+ |\lambda |).
\end{equation}
The
joint eigenvalue distribution which corresponds to (\ref{p(M)}) has the form (see \cite{Me:91})
\begin{equation}\label{p(la)}
p_{n,\beta}(\lambda_1,...,\lambda_n)=Q_{n,\beta}^{-1}\prod_{i=1}^n
e^{-n\beta V(\lambda_i)/2}\prod_{1\le i<j\le
n}|\lambda_i-\lambda_j|^\beta,
\end{equation}
where
\begin{equation}\label{Q}
Q_{n,\beta}=\int\prod_{i=1}^n
e^{-n\beta V(\lambda_i)/2}\prod_{1\le i<j\le
n}|\lambda_i-\lambda_j|^\beta d\lambda_1\dots d\lambda_n.
\end{equation}
  This distribution can be considered
for any $\beta>0$. We denote
\begin{equation}\label{E}
    \mathbf{E}_\beta\{(\dots)\}=\int(\dots)p_{n,\beta}(\lambda_1,...,\lambda_n)d\lambda_1,\dots
    d\lambda_n,
\end{equation}
and
\begin{equation}\label{p_nl}
p^{(n)}_{l,\beta}(\lambda_1,...,\lambda_l)=
\int_{\mathbb{R}^{n-l}} p_{n,\beta}(\lambda_1,...\lambda_l,\lambda_{l+1},...,\lambda_n)
d\lambda_{l+1}...d\lambda_n.
\end{equation}
It is known (see \cite{BPS:95,Jo:98}) that if $V'$ is a H\"{o}lder function, then
the first marginal density $p^{(n)}_{l,\beta}(\lambda)$ converges weakly to the density
$\rho(\lambda)$ (equilibrium density) with a compact support
$\sigma$. The support $\sigma$ and the density $\rho$ are uniquely defined by the conditions:
\begin{equation}\label{cond_rho}\begin{array}{l}\displaystyle
v(\lambda ):=2\int \log |\mu -\lambda |\rho (\mu )d\mu -V(\lambda )=\sup v(\lambda),\quad\lambda\in\sigma\\
v(\lambda )\le \sup v(\lambda),\quad \lambda\not\in\sigma,\hskip
2cm\sigma=\hbox{supp}\{\rho\}.
\end{array}\end{equation}
If we consider the linear eigenvalue statistics
of a smooth test  function $f$
\begin{equation}\label{lst}
    \mathcal{N}_n[f]=\sum_{i=1}^n f(\lambda_i),
\end{equation}
then the above results of \cite{BPS:95,Jo:98} mean that
\begin{eqnarray*}
  &&\lim_{n\to\infty}\mathbf{E}_\beta\left\{ n^{-1}
  \mathcal{N}_n[f]\right\}=\lim_{n\to\infty}\int f(\lambda)p^{(n)}_{l,\beta}(\lambda)d\lambda=
  \int\,f(\lambda)\rho(\lambda)d\lambda,\\
&&\lim_{n\to\infty}\mathbf{E}_\beta\left\{ |n^{-1} \mathcal{N}_n[f]-\mathbf{E}_\beta\{n^{-1}
\mathcal{N}_n[f]\}|^2\right\}=0.
\end{eqnarray*}

Moreover, in \cite{BPS:95} some rather rough bounds on the
rate of convergence  were found
\begin{eqnarray}\label{BPS:95}
   &&\bigg| \int f(\lambda)(p^{(n)}_{l,\beta}(\lambda)-\rho(\lambda))d\lambda\bigg|\le
  C||f||_2^{1/2}||f'||_2 ^{1/2}n^{-1/2}\log^{1/2}n,\\
  && \mathbf{E}_\beta\left\{ |n^{-1} \mathcal{N}_n[f]-\mathbf{E}_\beta\{n^{-1}
\mathcal{N}_n[f]\}|^2\right\}\le ||f||_2||f'||_2n^{-1}\log n.
\notag\end{eqnarray}
Here and below we denote by $||.||_2$ a standard $L^2(\sigma_\varepsilon)$- norm, with
$\sigma_\varepsilon$ being the $\varepsilon$-neighborhood of the support $\sigma$ with sufficiently
small $\varepsilon$.

In the case of $\beta=2$ these bounds can be improved considerably. It is a simple exercise
(see e.g. \cite{PS:97}) to show that for any $V$ satisfying (\ref{condV}) (not necessary Lipshitz)  the l.h.s. of
the second inequality is $\mathcal{O}(n^{-2})$, but for other $\beta$ this fact is not proven yet.
With the first inequality of (\ref{BPS:95}) the situation is similar. It follows from the results of \cite{DKMVZ:99}
that for real analytic $V$ the l.h.s. of the first inequality of (\ref{BPS:95}) is $\mathcal{O}(n^{-1})$
(see also \cite{APS:01} where the asymptotic expansion with respect to $n^{-1}$ was constructed
in the case of even real analytic $V$ and one or two interval support $\sigma$). Unfortunately,
similar results are not found for $\beta\not=2$ in the  general case of $\sigma$ till now.

Bounds of the type (\ref{BPS:95}) are interesting not only  themselves. They have a lot of
very important applications,  which includes  Central Limit Theorem
(CLT) for linear eigenvalue statistics, the asymptotic  for $\log Q_{n,\beta}$, etc. One of the most important and interesting applications is that
 to the universality problem for
$\beta=1,4$. Universality conjecture states
 that marginal densities (\ref{p_nl}) in the scaling limit, when
$\lambda_i=\lambda_0+x_i/n^\kappa$ $(i=1,\dots,l)$ are universal (i.e. they do not  depend on $V$).
The scaling exponent $\kappa$ depends on the behavior of the equilibrium density $\rho(\lambda)$
in a small neighborhood of $\lambda_0$. If $\rho(\lambda_0)\not=0$, then $\kappa=1$, if $\rho(\lambda_0)=0$
and $\rho(\lambda)\sim |\lambda-\lambda_0|^\alpha$, then $\kappa=1/(1+\alpha)$.

For $\beta=2$ universality of local eigenvalue statistics was proved in  many cases. For example, in the
 bulk case ($\rho(\lambda_0)\not=0$) it was shown in
\cite{PS:97} (see also \cite{PS:07}) that for a  general class of $V$ (the second
derivative of $V$ is Lipshitz in  some neighborhood of $\lambda_0$)  the scaled reproducing
kernel converges uniformly to the $\sin$-kernel. This result for the case of real analytic $V$ was obtained also in
\cite{DKMVZ:99}. Universality   in the bulk for very general conditions
on the potential $V$ was proved also recently in \cite{L-Lub:08}.
 Universality near the edge, i.e., the case when $\lambda_0$ is the edge point of
  the spectrum and $\rho(\lambda)\sim|\lambda-\lambda_0|^{1/2}$, as
$\lambda\sim\lambda_0$, was studied in \cite{DKMVZ:99}.
There are also results on universality near the extreme
point,  where $\rho(\lambda)\sim (\lambda-\lambda_0)^2$, as
$\lambda\sim\lambda_0$ (see \cite{C-K:06} for  real
analytic $V$ and \cite{S:05} for  general $V$).

The crucial difference between  the case $\beta=2$ and other $\beta$ is that for
 $\beta=2$ all correlation functions (\ref{p_nl}) can be expressed in terms of the reproducing
kernel of the system of  normalized  polynomials $p_j^{(n)}=\gamma_j^{(n)} x^j+\ldots$,
$(j=0,\dots,n-1)$  orthogonal on the real line with varying weight
\begin{equation}\label{w_n}
     w^{(n)} (\lambda):= e^{-n V(\lambda)}
\end{equation}
\begin{equation}\label{orth}
     \int_{\mathbb R} p_j^{(n)} (\lambda) p_k^{(n)} (\lambda) w^{(n)} (\lambda)\, d\lambda =\delta_{j,k}\quad
     \textrm{for}\ j,k \ge 0.
\end{equation}
The orthogonal polynomial machinery, in particular, Christoffel-Darboux formula and
Christoffel function simplify considerably the studies of  marginal densities (\ref{p_nl}).
Moreover, asymptotics of orthogonal polynomials $p_{n-1}^{(n)}$, $p_n^{(n)}$ are known
(see \cite{DKMVZ:99} for real analytic $V$ and the recent paper \cite{Mi-McL:08} for non analytic
$V$) and they can be used to prove bulk and edge universality.

For $\beta=1,4$ the situation is more complicated. It was shown in \cite{Tr-Wi:98} that
the problem  can be reduced to universality of some matrix kernels (see (\ref{Kn1}),
(\ref{Kn4}) below), which also can be expressed in terms of orthogonal polynomials (\ref{orth}),
but to control their behavior one need to control the invertibility of some matrix
 (see Section \ref{s:3} for more details).  According to Widom  \cite{Wi:99},
 if the potential $V$ is a rational function, then we need to control the inverse of some matrix
 of fixed size depending of $V$ (e.g., if $V$ is polynomial of degree $2m$, then we should
control some $(2m-1)\times(2m-1)$ matrix). Till now this technical problem was solved only in a few
cases. In the papers \cite{De-G:07,De-G:07a}  the case $V(\lambda)=\lambda^{2m}(1+o(1))$
(in  our notations) was studied. Similar method
  was used in \cite{DGKV:07} to prove bulk and edge universality (including the case
of hard edge) for the Laguerre type ensembles with monomial $V$.
 In \cite{St1} universality in the bulk and near the edges   were studied
for  $V$ being an even quatric polynomial.
In \cite{S:08,S:09} bulk and edge universality were studied  for $\beta=1$ and real
analytic even $V$  with one interval support $\sigma$.

But there is also a possibility to prove universality of local eigenvalue statistics by
using another technique.
In \cite{St1} Sojanovich made an important observation (see Remark 5 of \cite{St1}
or Section 3 of the present paper)
 which allows one to replace the problem to control  the Widom matrix by the problem to control
$\mathbf{E}_\beta\left\{ n^{-1} \mathcal{N}_n[f]\right\}$ for $\beta=1,2,4$. Thus the problem to study
the correction terms of the order $n^{-1}$ for $\mathbf{E}_\beta\left\{ n^{-1} \mathcal{N}_n[f]\right\}$
becomes especially important.

In a remarkable paper \cite{Jo:98} Johansson   studied
the expectation and the variance of $n^{-1}\mathcal{N}_n[f]$ up to the terms $\mathcal{O}(n^{-2})$.
This allows him, in particular, to prove CLT for  fluctuations of $\mathcal{N}_n[f]$.
Unfortunately, his method works only in the case of one interval support $\sigma$ of the equilibrium
density $\rho$ and polynomial $V$ with some additional assumption.

In the present paper we generalize the idea of \cite{Jo:98} to the case of real analytical $V$
with one interval support of $\rho$, without any other assumptions. Moreover, we give a more simple
proof of this result and  apply it  to the proof of bulk universality for $\beta=1,4$.

Let us formulate our main conditions.

\noindent \textbf{Condition C1.} \textit{The support $\sigma $ of the equilibrium measure
 density $\rho$ consists of a single interval:}
$\sigma =[a,b]$, $-\infty <a<b<\infty.$

\begin{remark} It is easy to see that changing the variables
 $M^{\prime }=2(M-\displaystyle\frac{a+b}{2}I)/(b-a)$, in the case (i) we can
always take the support
$\sigma =[-2,2]$.
\end{remark}
\medskip \noindent \textbf{Condition C2. }\textit{ The equilibrium density $\rho$ can be represented in the form
\begin{equation}\label{rho}
    \rho(\lambda)=\frac{1}{2\pi}P(\lambda)\Im X^{1/2}(\lambda+i0),\quad
    \inf_{\lambda\in[-2,2]}P(\lambda)>0,
\end{equation}
where
\begin{equation}\label{X}
    X(z)= z^2-4,
    \end{equation}
and we choose a branch of $X^{1/2}(z)$ such that $X^{1/2}(z)\sim z$, as $z\to+\infty$.
Moreover, the function $v$ defined by (\ref{cond_rho})
attains its maximum if and only if $\lambda $ belongs to  $\sigma $. }

\medskip \noindent \textbf{Condition C3. }\textit{$V$ is real
analytic on $\sigma $, i.e., there exists an open domain $\mathbf{D}\subset
\mathbb{C}$ such that $\sigma\subset\mathbf{D}$ and  ${V}$  is an analytic function in $\mathbf{D}$.}

\begin{remark} It is known (see, e.g., \cite{APS:01}) that under conditions C1 and C3 for any
$\beta$ the equilibrium density $\rho$
of the ensemble (\ref{p(la)}) has the form (\ref{rho}) -- (\ref{X}) with $P\ge 0$.
 The analytic in $\mathbf{D}$  function $P$ in
(\ref{rho})  can be represented in the form
\begin{equation}\label{P}
    P(z)=\int_\sigma\frac{V'(z)-V'(\lambda)}{(z-\lambda)\Im X^{1/2}(\lambda+i0)}d\lambda
    \end{equation}
Hence, condition C2 states that $P$ has no zeros in $[-2,2]$. Note also, that in the paper
\cite{Jo:98} it was assumed additionally that $V$ is a polynomial and $P$ has no zeros on the real line.
\end{remark}

The first result of the paper is the theorem which allows us to control
the expectation and the variance of linear eigenvalue statistics.

\begin{theorem}\label{t:1}
Under  conditions C1 -- C3 for any analytic in $\mathbf{D}$ function $f$ we have
\begin{multline}\label{t1.1}
\mathbf{E}_\beta\{\mathcal{N}_n[f]\}=\int f(\lambda)\rho(\lambda)d\lambda\\+
\frac{1}{n}\left(\frac{2}{\beta}-1\right)\frac{1}{(2\pi i)^2 }\oint_{\mathcal{L}_{2d}}
 \frac{f(z)dz}{X^{1/2}(z)}\oint_{\mathcal{L}_{d}}
\frac{g'(\zeta)d\zeta}{P(\zeta)(z-\zeta)}+n^{-2}r_{n,\beta}(f),
\end{multline}
where the contour $\mathcal{L}_{d}$ is defined as
\begin{equation}\label{L_d}
\mathcal{L}_{d}=\{z:\hbox{dist}\{z,\sigma\}=d\},
\end{equation}
 $d$ is chosen sufficiently small  to have all zeros of $P(\zeta)$ outside of
$\mathcal{L}_{2d}$,
\begin{equation}\label{g}
  g(z)=  \int\frac{\rho(\lambda)d\lambda
    }{z-\lambda},
\end{equation}
and $r_{n,\beta}(f)$ satisfies the bound
\[|r_{n,\beta}(f)|\le C_d\sup_{z:\emph{dist}\{z,\sigma\}\le 2d}|f(z)|,\]
with $C_d$ depending only on $d$.

Moreover,
\begin{equation}\label{t1.2}
\mathbf{E}_\beta\left\{ | \mathcal{N}_n[f]-\mathbf{E}_\beta\{
\mathcal{N}_n[f]\}|^2\right\}
\le C_d\sup_{z:\emph{dist}\{z,\sigma\}\le 2d}|f(z)|^2.
\end{equation}
\end{theorem}

 One of the important applications of Theorem \ref{t:1} (see discussion above) is the asymptotic
of $\log Q_{n,\beta}$. Since the paper \cite{BPS:95} it is known that
\[n^{-2}\log Q_{n,\beta}=\frac{\beta}{2}\mathcal{E}_V+\mathcal{O}(\log n/n),\]
where
\begin{equation}\label{E_V}
\mathcal{E}_V=-\int\log\frac{1}{|\lambda-\mu|}\rho(\lambda)\rho(\mu)d\lambda d\mu-
\int V(\lambda)\rho(\lambda)d\lambda.
\end{equation}
But for many problems it is important to control the next terms of asymptotic expansion
of $\log Q_{n,\beta}$ (for applications see  discussion in \cite{Er-McL:02}, where the complete
asymptotic expansion with respect to $n^{-1}$ was constructed for the case $\beta=2$ under assumption that $V$
is a polynomial close in a certain sense  to $V_0(\lambda)=\lambda^2/2$.)
\begin{theorem}\label{t:log}
Under  conditions C1 -- C3 for any $\beta$
\begin{eqnarray}\label{log}
n^{-2}\log Q_{n,\beta}&=&n^{-2}\log Q_{n,\beta}^{(0)}+\frac{1}{2}\beta\mathcal{E}_V+\frac{3}{8}\beta
\\
&&+\frac{1}{n}\left(1-\frac{\beta}{2}\right)\frac{1}{(2\pi i)^2 }\oint_{\mathcal{L}_{2d}}
 \frac{(V(z)-z^2/2)dz}{X^{1/2}(z)}\oint_{\mathcal{L}_{d}}
\frac{g'_t(\zeta)d\zeta}{P_t(\zeta)(z-\zeta)}+\mathcal{O}(n^{-2}),
\notag\end{eqnarray}
where $\log Q_{n,\beta}^{(0)}$ corresponds to the Gaussian case $V_0=\lambda^2/2$,
$\mathcal{E}_V$ is defined by (\ref{E_V}),  $\frac{3}{8}\beta=-\frac{1}{2}\beta\mathcal{E}_{V_0}$,
and
\[P_t(\lambda)=tP(\lambda)+1-t,\quad g_t(z)=tg(z)+\frac{1-t}{2}(z-\sqrt{z^2-4}).\]
\end{theorem}
\begin{remark}\label{r:Sielb} By the Selberg formula (see  e.g. \cite{Me:91})  for the Gaussian case
we have
\[
Q_{n,\beta}^{(0)}=n!\left(\frac{n\beta}{2}\right)^{-\beta n(n-1)/4-n/2}(2\pi)^{n/2}\prod_{j=1}^{n}
\frac{\Gamma(\beta j/2)}{\Gamma(\beta /2)}\]
\end{remark}
As it was mentioned above, Theorem \ref{t:1} together with some
asymptotic results of \cite{DKMVZ:99} for orthogonal polynomials  can be used  to prove universality
of the local eigenvalue statistics of the matrix models (\ref {p(M)}). We  restrict our
attention to the case when $V$ is a polynomial
of even degree $2m$ such that conditions C1--C3 are satisfied. Moreover we consider
only  even $n$.
It is  known (see \cite{Tr-Wi:98}) that the question of universality is closely
related to the large $n$ behavior of certain matrix kernels
\begin{align}\label{Kn1}
     K_{n,1} (\lambda,\mu)&:=
     \begin{pmatrix}
          S_{n,1}(\lambda,\mu) & -\frac{\partial}{\partial \mu} S_{n,1}(\lambda,\mu)\\
          (\epsilon S_{n,1}) (\lambda,\mu)-\epsilon (\lambda-\mu) & S_{n,1}(\mu,\lambda)
     \end{pmatrix}\ \textrm{for}\ \beta=1, n\ \textrm{even,}\\
     \label{Kn4}
     K_{n,4}(\lambda,\mu) &:=
     \begin{pmatrix}
          S_{n,4}(\lambda,\mu) & -\frac{\partial}{\partial \mu} S_{n,4}(\lambda,\mu)\\
          (\epsilon S_{n,4}) (\lambda,\mu) & S_{n,4}(\mu,\lambda)
     \end{pmatrix}\ \textrm{for}\ \beta=4.
\end{align}
Here $\epsilon (\lambda)=\frac{1}{2} \hbox{sgn} (\lambda)$, where $\hbox{sgn}$ denotes the standard signum
function, and
$(\epsilon S_{n,\beta})(\lambda,\mu)=\int_{\mathbb R}\epsilon (x-x')S_{n,\beta}(x',y)\, d\lambda'$.
Some formulae for the functions $S_{n,\beta}$ that appear in the definition of
$K_{n,\beta}$ will be introduced in (\ref{Sn1}), (\ref{Sn4}) below. In order
to state our theorem we need some more notation.
Define
\begin{align*}
     K_\infty (t)&:= \frac{\sin \pi t}{\pi t},\\
     K_\infty^{(1)} (\xi, \eta)&:=
     \begin{pmatrix}
          K_\infty (\xi-\eta) & K'_\infty (\xi-\eta)\\
          \int^{\xi-\eta}_0 K_\infty (t)\, dt -\epsilon (\xi-\eta) & K_\infty (\eta-\xi )
     \end{pmatrix},\\
     K_\infty^{(4)} (\xi, \eta)&:=
     \begin{pmatrix}
          K_\infty (\xi-\eta) & K'_\infty (\xi-\eta)\\
          \int^{\xi-\eta}_0 K_\infty (t)\, dt & K_\infty (\eta-\xi )
     \end{pmatrix}.
\end{align*}
Furthermore we denote for a $2\times 2$ matrix $A$ and $\lambda >0$
\begin{equation*}
      A^{(\lambda)} := \begin{pmatrix}
           \sqrt{\lambda}^{-1} & 0\\
           0&\sqrt{\lambda}
      \end{pmatrix}
      A \begin{pmatrix}
           \sqrt{\lambda} & 0\\
           0&\sqrt{\lambda}^{-1}
      \end{pmatrix}.
\end{equation*}

\begin{theorem} \label{Tt:2}
Let $V$ be a polynomial of degree $2m$ with positive leading coefficient and such that
conditions C1--C2 are satisfied. Then we have for (even) $n\to \infty$,
$\lambda_0\in\mathbb R$ with $\rho (\lambda_0)>0$, and for $\beta \in \{1,4\}$ that
\begin{align*}
     &\frac{1}{q_n} K_{n,1}^{(q_n)} \left(\lambda_0+\frac{\xi}{q_n},\lambda_0+\frac{\eta}{q_n} \right)
     =K_\infty^{(1)} (\xi,\eta)+\mathcal O(n^{-1/2}),\\
     &\frac{1}{q_n} K_{n/2,4}^{(q_n)} \left(\lambda_0+\frac{\xi}{q_n},\lambda_0+\frac{\eta}{q_n} \right)
     =K_\infty^{(4)} (\xi,\eta)+\mathcal O(n^{-1/2}),
\end{align*}
where $q_n=n\rho(\lambda_0)$. The error bound is uniform for bounded $\xi$, $\eta$ and for
$\lambda_0$ contained in some compact subset of $(-2,2)$ (recall that $\hbox{supp}\, \rho =[-2,2]$
by Condition C1).
\end{theorem}

It is an immediate consequence of Theorem \ref{Tt:2} that the corresponding rescaled $l$-point
correlation functions
\begin{equation*}
     p_{l,1}^{(n)} \left(\lambda_0+\frac{\xi_1}{q_n},\ldots ,\lambda_0+\frac{\xi_l}{q_n}\right),\quad
     p_{l,4}^{(n/2)} \left(\lambda_0+\frac{\xi_1}{q_n},\ldots ,\lambda_0+\frac{\xi_l}{q_n}\right)
\end{equation*}
converge for $n$ (even) $\to \infty$ to some limit that depends
on $\beta$ but not on the choice of $V$.

The paper is organized as follows. In Section \ref{s:2} we prove Theorems \ref{t:1} and
\ref{t:log}. In Section \ref{s:3} we prove Theorem \ref{Tt:2} modulo some  bounds, which we
obtain in Section \ref{s:ap1}. And in Section \ref{s:ap2} for the reader's convenience  we give
a version of the proof of  a priory bound (\ref{BPS:95}).

\section{Proof of Theorems \ref{t:1}, \ref{t:log}}\label{s:2}
\textbf{Proof of Theorem \ref{t:1}}. Take $n$-independent $\varepsilon$, small enough to
provide that $\sigma_\varepsilon\subset\mathbf{D}$, where $\sigma_\varepsilon\subset\mathbb{R}$ means the
$\varepsilon$-neighborhood
of $\sigma$. It is known (see e.g. \cite{PS:07}))
that if we replace in (\ref{p(la)}),(\ref{E}) and (\ref{p_nl}) the integration over $\mathbb{R}$
by the integration $\sigma_\varepsilon$, then the new  marginal densities will differ from
the initial ones by the terms  $\mathcal{O}(e^{-nc})$ with some $c$ depending on $\varepsilon$, but
independent of $n$. Since for our purposes it is more convenient to consider the
integration with respect to $\sigma_\varepsilon$, we assume from this moment that this
replacement is made, so everywhere below the integration without limits means the
integration over $\sigma_\varepsilon$.

Following  the idea of
\cite{Jo:98}, we will study a little bit modified form of the joint eigenvalue
distribution, than in (\ref{p(la)}). Namely, consider any real on $\sigma$ and
analytic in $\mathbf{D}$ function $h(\zeta)$ and denote
\[V_h(\zeta)=V(\zeta)+\frac{1}{n}h(\zeta).\]
Let $p_{n,\beta,h}$, $\mathbf{E}_{\beta,h}\{\dots\}$, $p^{(n)}_{l,\beta,h}$  be the distribution
density, the expectation, and the marginal densities defined by (\ref{p(la)}),(\ref{E}) and (\ref{p_nl})
with $V$ replaced by $V_h$.

By (\ref{p(la)})  the first  marginal density can be represented in the
form
\begin{equation}\label{r_rho}
p^{(n)}_{1,\beta,h}(\lambda)=Q_{n,\beta,h}^{-1}\int
e^{-n\beta V_h(\lambda)/2}\prod_{i=2}^n|\lambda-\lambda_i|^\beta
e^{-n\beta V_h(\lambda_i)/2}\prod_{2\le i<j\le
n}|\lambda_i-\lambda_j|^\beta d\lambda_2\dots d\lambda_n.
\end{equation}
Using the representation and integrating by parts,
we obtain
\begin{equation}\label{eq_1}
\int\frac{V'_h(\lambda)p^{(n)}_{1,\beta,h}(\lambda)}{z-\lambda} d\lambda=
\frac{2}{\beta n}\int\frac{p^{(n)}_{1,\beta,h}(\lambda)}{(z-\lambda)^2} d\lambda+
\frac{2(n-1)}{n}\int\frac{p^{(n)}_{2,\beta,h}(\lambda,\mu)d\lambda
d\mu}{(z-\lambda)(\lambda-\mu)}+\mathcal{O}(e^{-nc}).
\end{equation}
Here  $\mathcal{O}(e^{-nc})$ is the contribution of the integrated term. In fact all equations
below should contain $\mathcal{O}(e^{-nc})$, but in order to simplify formula below we  omit it.

Since the function $p^{(n)}_{2,\beta,h}(\lambda,\mu)$ is symmetric with respect to
$\lambda,\mu$, we have
\[2\int\frac{p^{(n)}_{2,\beta,h}(\lambda,\mu)d\lambda d\mu}{(z-\lambda)(\lambda-\mu)}=
\int\frac{p^{(n)}_{2,\beta,h}(\lambda,\mu)d\lambda d\mu}{(z-\lambda)(\lambda-\mu)}+
\int\frac{p^{(n)}_{2,\beta,h}(\lambda,\mu)d\lambda d\mu}{(z-\mu)(\mu-\lambda)}=
\int\frac{p^{(n)}_{2,\beta,h}(\lambda,\mu)d\lambda d\mu}{(z-\lambda)(z-\mu)}.
\]
Hence, equation (\ref{eq_1}) can be written in the form
\begin{equation}\label{eq_2}
\int\frac{V'_h(\lambda)p^{(n)}_{1,\beta,h}(\lambda)}{z-\lambda} d\lambda=
\frac{2}{\beta n}\int\frac{p^{(n)}_{1,\beta,h}(\lambda)}{(z-\lambda)^2} d\lambda+
\frac{(n-1)}{n}\int\frac{p^{(n)}_{2,\beta,h}(\lambda,\mu)d\lambda
d\mu}{(z-\lambda)(z-\mu)}.
\end{equation}
Let us introduce notations:
\begin{multline}\label{delta}
    \delta_{n,\beta,h}(z)=n(n-1)\int\frac{p^{(n)}_{2,\beta,h}(\lambda,\mu)d\lambda
    d\mu}{(z-\lambda)(z-\mu)}-n^2\bigg(\int\frac{p^{(n)}_{1,\beta,h}(\lambda)d\lambda
    }{z-\lambda}\bigg)^2+n\int\frac{p^{(n)}_{1,\beta,h}(\lambda)}{(z-\lambda)^2}
    d\lambda\\=\int \frac{k_{n,\beta,h}(\lambda,\mu)d\lambda d\mu}{(z-\lambda)(z-\mu)},\hskip3cm
\end{multline}
where
\begin{equation}\label{k_n}
k_{n,\beta,h}(\lambda,\mu)=n(n-1)p^{(n)}_{2,\beta,h}(\lambda,\mu)-n^2p^{(n)}_{1,\beta,h}(\lambda)
p^{(n)}_{1,\beta,h}(\mu)+n\delta(\lambda-\mu)p^{(n)}_{1,\beta,h}(\lambda).
\end{equation}
Moreover, we denote
\begin{equation}\label{g_n}
  g_{n,\beta,h}(z)=  \int\frac{p^{(n)}_{1,\beta,h}(\lambda)d\lambda
    }{z-\lambda},\quad V(z,\lambda)=\frac{V'(z)-V'(\lambda)}{z-\lambda}.
\end{equation}
Then equation (\ref{eq_1}) takes the form
\begin{multline}\label{eq_3}
g_{n,\beta,h}^2(z)-V'(z)g_{n,\beta,h}(z)+\int V(z,\lambda)p^{(n)}_{1,\beta,h}(\lambda) d\lambda\\
=\frac{1}{n}\int\frac{h'(\lambda)p^{(n)}_{1,\beta,h}(\lambda)}{z-\lambda} d\lambda-
\frac{1}{n}\bigg(\frac{2}{\beta }-1\bigg)\int\frac{p^{(n)}_{1,\beta,h}(\lambda)}{(z-\lambda)^2} d\lambda
-\frac{1}{n^2}\delta_{n,\beta,h}(z).
\end{multline}
Using that $V(z,\zeta)$ is an analytic function of $\zeta$ in $\mathbf{D}$, we obtain by the
Cauchy  theorem that for any $z$ outside of $\mathcal{L}_d$
\begin{eqnarray*}
\int V(z,\lambda)p^{(n)}_{1,\beta,h}(\lambda) d\lambda=\frac{1}{2\pi i}\oint_{\mathcal{L}_d}
V(z,\zeta)g_{n,\beta,h}(\zeta)d\zeta,
%\\
%\int\frac{h'(\lambda)p^{(n)}_{1,\beta,h}(\lambda)}{z-\lambda} d\lambda=
%\frac{1}{2\pi i n}\oint_{\mathcal{L}_d}\frac{h'(\zeta)g_{n,\beta,h}(\zeta)}{z-\zeta}d\zeta
\end{eqnarray*}
Thus, (\ref{eq_3}) takes the form
\begin{multline}\label{eq_4}
g_{n,\beta,h}^2(z)-V'(z)g_{n,\beta,h}(z)+\frac{1}{2\pi i}\oint_{\mathcal{L}_d}
V(z,\zeta)g_{n,\beta,h}(\zeta)d\zeta\\=
\frac{1}{n}\int\frac{h'(\lambda)p^{(n)}_{1,\beta,h}(\lambda)}{z-\lambda}d\lambda-
\frac{1}{n}\bigg(\frac{2}{\beta }-1\bigg)\int\frac{p^{(n)}_{1,\beta,h}(\lambda)}
{(z-\lambda)^2}
d\lambda-\frac{1}{n^2}\delta_{n,\beta,h}(z).
\end{multline}
Passing to the limit $n\to\infty$, we obtain for any fixed $z$ the quadratic equation
\begin{equation}\label{eq_5}
g^2(z)-V'(z)g(z)+Q(z)=0,\quad Q(z)=\frac{1}{2\pi i}\oint_{\mathcal{L}_d}
V(z,\zeta)g(\zeta)d\zeta,
\end{equation}
where $g$ is defined by (\ref{g}).
Hence,
\[
g(z)=\frac{1}{2}V'(z)-\frac{1}{2}\sqrt{V'(z)^2-4Q(z)}.
\]
Using the inverse Stieltjes transform and comparing with (\ref{rho}), we get that
\begin{equation}\label{2g-V}
2g(z)-V'(z)=P(z) X^{1/2}(z).\end{equation}
where $X(z)$ is defined by (\ref{X}).

Denote
\begin{equation}\label{u_n}
  u_{n,\beta,h}(z)=  n(g_{n,\beta,h}(z)-g(z))\quad\Leftrightarrow\quad
  g_{n,\beta,h}(z)=g(z)+\frac{1}{n}u_{n,\beta,h}(z).
\end{equation}
Then, subtracting (\ref{eq_5}) from (\ref{eq_4}) and multiplying the result by $n$, we get
\begin{equation}\label{eq_6}
(2g(z)-V'(z))u_{n,\beta,h}(z)+\frac{1}{2\pi i}\oint
V(z,\zeta)u_{n,\beta,h}(\zeta)d\zeta=F(z),
\end{equation}
where
\begin{eqnarray}\label{F}
F(z)&=&
\int\frac{h'(\lambda)p^{(n)}_{1,\beta,h}(\lambda)}{z-\lambda}d\lambda
+\bigg(\frac{2}{\beta }-1\bigg)\left(g'(z)+\frac{1}{n}u_{n,\beta,h}'(z)\right)\\
&&-\frac{1}{n}u_{n,\beta,h}^2(z)-\frac{1}{n}\delta_{n,\beta,h}(z).
\notag\end{eqnarray}
Using (\ref{2g-V}), we obtain from (\ref{eq_6})
\begin{equation}\label{eq_7}
P(z)X^{1/2}(z)u_{n,\beta,h}(z)+\mathcal{Q}_n(z)=F(z),\quad
\mathcal{Q}_n(z)=\frac{1}{2\pi i}\oint
V(z,\zeta)u_{n,\beta,h}(\zeta)d\zeta.
\end{equation}
Then, choosing $d$ such that the contour $\mathcal{L}_d$ defined by (\ref{L_d})
does not contain
 zeros of $P(\zeta)$, we
get for any $z$  outside of $\mathcal{L}_d$
\begin{equation}\label{eq_8}
\frac{1}{2\pi i}\oint_{\mathcal{L}_d}\left(P(\zeta)X^{1/2}(\zeta)u_{n,\beta,h}(\zeta)
+\mathcal{Q}_n(\zeta)-F(\zeta)\right)\frac{d\zeta}{P(\zeta)(z-\zeta)}=0.
\end{equation}
Since,  by definition (\ref{eq_7}), $\mathcal{Q}_n(\zeta)$ is an analytic function in
$\mathbf{D}$, and $z$ and all zeros of $P$ are outside of $\mathcal{L}_d$, the Cauchy
theorem yields
\[
\frac{1}{2\pi i}\oint_{\mathcal{L}_d}\frac{\mathcal{Q}_n(\zeta)d\zeta}{P(\zeta)(z-\zeta)}=0.
\]
Moreover,   since
\[u_{n,\beta,h}(z)=\frac{n}{z}\left(\int d\lambda p^{(n)}_{1,\beta,h}(\lambda)-
\int d\lambda \rho(\lambda)\right)+n\mathcal{O}(z^{-2})=n\mathcal{O}(z^{-2})
,\quad z\to \infty
\]
we have
\begin{equation}\label{as_u}
X^{1/2}(z)u_{n,\beta,h}(z)=n\mathcal{O}(z^{-1}).
\end{equation}
Then the Cauchy theorem yields
\[\frac{1}{2\pi
i}\oint_{\mathcal{L}}\frac{X^{1/2}(\zeta)u_{n,\beta,h}(\zeta)d\zeta}{(z-\zeta)}=X^{1/2}(z)u_{n,\beta,h}(z).
\]
Finally, we obtain from (\ref{eq_8})
\begin{equation}\label{eq_9}
u_{n,\beta,h}(z)=\frac{1}{2\pi iX^{1/2}(z)}\oint_{\mathcal{L}_d} \frac{F(\zeta)d\zeta}{P(\zeta)(z-\zeta)}.
\end{equation}
Now take $d$ small enough to have all zeros of $P$ outside of $\mathcal{L}_{3d}$.
Then for any $z:\hbox{dist}\{z,\sigma\}=2d$ equation (\ref{eq_9}) implies
\begin{equation}\label{eq_10}
u_{n,\beta,h}(z)=\frac{F(z)}{X^{1/2}(z)P(z)}+
\frac{1}{2\pi iX^{1/2}(z)}\oint_{\mathcal{L}_{3d} }\frac{F(\zeta)d\zeta}{P(\zeta)(\zeta-z)}.
\end{equation}
According to the result of \cite{BPS:95} for any $\beta$ we have a priory bound
\begin{equation}\label{b1_de}
    |\delta_{n,\beta,h}|\le \frac{C n\log n}{\hbox{dist}^4\{z,\sigma\}},\quad
    \quad |u_{n,\beta,h}(z)|\le\frac{C n^{1/2}\log n}{\hbox{dist}^2\{z,\sigma\}},
     \quad |u_{n,\beta,h}'(z)|\le \frac{C n^{1/2}\log n}{\hbox{dist}^3\{z,\sigma\}},
\end{equation}
where $C$ is an absolute constant.

 Denote
\[ M_n(d)=\sup_{z:\hbox{dist}\{z,\sigma\}\ge 2d}|u_{n,\beta,h}(z)|\]
By (\ref{as_u}) and the maximum principle, there exists a point
$z:\hbox{dist}\{z,\sigma\}=2d$ such that
\[
M_n(d)=|u_{n,\beta,h}(z)|.
\]
Then, using (\ref{eq_10}), the definition of $F$ (see (\ref{F})), and (\ref{b1_de}),
we obtain the inequality
\[M_n(d)\le \frac{1}{n}C_1M_n^2(d)+C_2\log n,\]
where $C_1$ and $C_2$ depend only on $d$, $\displaystyle\sup_{\hbox{\small dist}\{z,\sigma\}\le
3d}|P^{-1}(z)|$, $\displaystyle\sup_{\hbox{\small dist}\{z,\sigma\}\le
d/2}|n^{-1}h(z)|$,  and from $C$ of (\ref{b1_de}).
Solving the above  quadratic inequality, we get
\[
\left[\begin{array}{l}
M_n(d)\ge (2C_1)^{-1}(n+\sqrt{n^2-4C_1C_2n\log n})\\
M_n(d)\le(2C_1)^{-1}(n-\sqrt{n^2-4C_1C_2n\log n})
\end{array}\right.
\]
Since the first inequality contradicts to  (\ref{b1_de}), we conclude that the second
inequality holds. Hence, we get
\[
\sup_{z:\hbox{dist}\{z,\sigma\}\ge 2d}|u_{n,\beta,h}(z)|\le 2C_2\log n+C(\sup_
{\lambda\in\sigma_\varepsilon}|h'(\lambda)|+\hbox{dist}^{-2}\{z,\sigma\}).
\]
Note that the bound gives us that for any real analytic
$\varphi(\zeta)$
\begin{eqnarray}\label{b1_u}
n\bigg|\int\varphi(\lambda)(p^{(n)}_{1,\beta,h}(\lambda)-\rho(\lambda))d\lambda\bigg|&=&
\bigg|\frac{1}{2\pi i}\oint_{\mathcal{L}_{2d}
}\varphi(\zeta)u_{n,\beta,h}(\zeta)d\zeta\bigg|\\&\le&
w_n \left(\sup_{z\in\mathcal{L}_{2d}}|\varphi(z)|+\sup_{\lambda\in\sigma_\varepsilon}
|h'(\lambda)|\right),
\notag\end{eqnarray}
where
\[w_n =2C_2\log n\]
 Now we are going to use the following lemma, which is an analog of Lemma 3.11 of
 \cite{Jo:98}.
 \begin{lemma}\label{l:1}
 If (\ref{b1_u}) holds for any real  $h$, and some $\varphi$ which is analytic in
 $\mathbf{D_1}\subset\mathbf{D}$ ($\sigma_\varepsilon\subset\mathbf{D_1}$), then there exists
 an $n$-independent constant $C_*$ such
 that
\begin{equation}\label{bl_de}
\int k_{n,\beta,h}(\lambda,\mu)\varphi(\lambda)\varphi(\mu)d\lambda d\mu\le C_*w_n^2\sup|\varphi^2|
\end{equation}
 \end{lemma}
 The lemma was proved in \cite{Jo:98}, but for  convenience of readers we
 give its proof
at the end of the proof of Theorem \ref{t:1}.

 Applying the lemma to $\varphi_z^{(1)}(\lambda)=\Re(z-\lambda)$ and
 $\varphi_z^{(2)}(\lambda)=\Im(z-\lambda)$ with $\hbox{dist}\{z,\sigma\}\ge d$, and using
 (\ref{b1_u}),  we obtain  for such $z$ (cf (\ref{b1_de}))
\begin{equation}\label{b2_de}
    |\delta_{n,\beta,h}|\le C_d'\log^2 n,\quad |u_{n,\beta,h}(z)|,|u_{n,\beta,h}'(z)|\le C_d'\log n.
\end{equation}
Then, using this bound in (\ref{eq_10}) instead of (\ref{b1_de}), by the same way as above
 we get (\ref{b1_u}) with $w_n =C_1(\sup_{\lambda\in\sigma_\varepsilon}|h'(\lambda)|+C_d)$.
  Then, applying Lemma \ref{l:1} once more, we obtain
 that
\begin{equation}\label{b_de}
    |\delta_{n,\beta,h}|\le C_d'',
    \quad |u_{n,\beta,h}(z)|,\,|u_{n,\beta,h}'(z)|\le C_d''.
\end{equation}
Using these final bounds in (\ref{eq_9}), we obtain that
\begin{equation}\label{eq_11}
u_{n,\beta,h}(z)=\frac{1}{2\pi iX^{1/2}(z)}\oint_{\mathcal{L}_d} \frac{g'(\zeta)d\zeta}{P(\zeta)(z-\zeta)}+
r_n(z),
\end{equation}
where
\[|r_n(z)|\le n^{-1}C_d.\]
$\square$

\medskip

\textbf{Proof of Lemma \ref{l:1}}. Take any real analytic $\varphi$ such that
$\sup_{z\in\mathcal{L}_{2d}}|\varphi(z)|\le 1$. Using the method of \cite{Jo:98},
consider the function
\[ F_n(t)=\mathbf{E}_{\beta,h}\left\{\exp\left[\frac{t}{2w_n}\sum_{i=1}^n(\varphi(\lambda_i)
-\int\varphi(\lambda)\rho(\lambda)d\lambda)\right]\right\}.
\]
It is easy to see that
\begin{equation}\label{logF}
\frac{d^2}{dt^2}\log F_n(t)=(2w_n)^{-2}\mathbf{E}_{\beta,h+t\varphi/2w_n}
\left\{\left(\sum_{i=1}^n(\varphi(\lambda_i)-\mathbf{E}_{\beta,h+t\varphi/2w_n}\{\varphi(\lambda_i)\})
\right)^2\right\}\ge 0.
\end{equation}
Hence, by (\ref{b1_u}), for $t\in[-1,1]$
\begin{multline*}
\log F_n(t)=\log F_n(t)-\log F_n(0)=\int_0^t\frac{d}{d\tau}\log F_n(\tau)d\tau\le |t|\frac{d}{dt}\log
F_n(t)\\=|t|(2w_n)^{-1}\mathbf{E}_{\beta,h+t\varphi/2w_n}
\left\{\sum_{i=1}^n\left(\varphi(\lambda_i)-\int\varphi(\lambda)\rho(\lambda)d\lambda\right)\right\}\\
=\frac{|t|n}{2w_n}\int\varphi(\lambda)\left(p^{(n)}_{1,\beta,h+t\varphi/2w_n}(\lambda)-\rho(\lambda)\right)d\lambda\le |t|
\end{multline*}
Thus, for $t\in[-1,1]$
\[F_n(t)\le e^{|t|}\le 3,\]
and for any $t\in \mathbb{C}$, $|t|\le 1$
\begin{equation}\label{b_F}
   |F_n(t)| \le F_n(|t|)<3.
\end{equation}
Then, we have by the Cauchy theorem, for $|t|\le \frac{1}{2}$
\[ |F_n'(t)|=\bigg|\frac{1}{2\pi}\oint_{|t'|=1}\frac{F_n(t')dt'}{(t'-t)^2}\bigg|\le 6,\]
and therefore for $|t|\le\frac{1}{15}$
\[ |F_n(t)|= \Big|F(0)-\int_{0}^tF_n'(t)dt\Big|\ge \frac{1}{2}.\]
Hence, $\log F_n(t)$ is an analytic function for $|t|\le\frac{1}{12}$ and so,
using the above bounds, we have
\[
\frac{d^2}{dt^2}\log F_n(0)=
\frac{1}{2\pi i}\oint_{|t|=1/12}\frac{\log F_n(t)}{t^3}dt\le C.
\]
Finally, using (\ref{logF}), we get
\[\int k_{n,\beta,h}(\lambda,\mu)\varphi(\lambda)\varphi(\mu)d\lambda d\mu=
\mathbf{E}_{\beta,h}
\left\{\left(\sum_{i=1}^n(\varphi(\lambda_i)-\mathbf{E}_{\beta,h}\{\varphi(\lambda_i)\})\right)^2\right\}
\le 4Cw_n^2.\]
$\square$
\medskip

\textbf{Proof of Theorem \ref{t:log}}
Consider  the functions $V_t$ of the form
\begin{equation}\label{V_t}
    V_t(\lambda)=tV(\lambda)+(1-t)V_0(\lambda),
\end{equation}
where $V_0(\lambda)=\lambda^2/2$.
Let $Q_{n,\beta}(t)$ be defined by (\ref{Q})
with $V$ replaced by $V_t$. Then evidently $Q_{n,\beta}(1)=Q_{n,\beta}$ and $Q_{n,\beta}(0)$
 corresponds to the Gaussian case $V_0(\lambda)=\lambda^2/2$. Hence,
\begin{align}\label{d_log_Q}
\frac{1}{n^2}\log Q_{n,\beta}(1)-\frac{1}{n^2}\log Q_{n,\beta}(0)&=
\frac{1}{n^2}\int_0^1dt\frac{d}{dt}\log Q_{n,\beta}(t)\\
&=\frac{\beta}{2}\int_0^1dt\int d\lambda(V(\lambda)-V_0(\lambda))p^{(n)}_{1,\beta}(\lambda;t),\notag
\end{align}
where $p^{(n)}_{1,\beta}(\lambda;t)$ is the first marginal density corresponding to $V_t$.
Using (\ref{cond_rho}) one can check that  if we consider the distribution
(\ref{p(la)}) with $V$ replaced by $V_t$, then the limiting DOS $\rho_t$ has the form
\begin{equation}\label{rho_t}
\rho_t(\lambda)=t\rho(\lambda)+(1-t)\rho_0(\lambda)=\frac{1}{2\pi}\sqrt{4-\lambda^2}
\left[tP(\lambda)+(1-t)P_0(\lambda)\right],
\end{equation}
with $X$ defined by (\ref{X}) and $P_0(\lambda)=1$. Hence, using (\ref{t1.1})
for the last integral in (\ref{d_log_Q}), we get
\begin{eqnarray*}
\frac{1}{n^2}\log Q_{n,\beta}&=&\frac{1}{n^2}\log Q_{n,\beta}(0)
-\frac{\beta}{2}\mathcal{E}_0+\frac{\beta}{2}\mathcal{E}_V\\
&&+\frac{1}{n}\left(1-\frac{\beta}{2}\right)\frac{1}{(2\pi i)^2 }\oint_{\mathcal{L}_{2d}}
 \frac{(V(z)-z^2/2)dz}{X^{1/2}(z)}\oint_{\mathcal{L}_{d}}
\frac{g'_t(\zeta)d\zeta}{P_t(\zeta)(z-\zeta)}+\mathcal{O}(n^{-2}),
\end{eqnarray*}
where $\mathcal{E}_V$ is defined by (\ref{E_V}), $P_t$ and $g_t$ are defined in (\ref{log}) and
$\mathcal{E}_0=-\frac{3}{4}$ (see, e.g., \cite{Me:91}).
$\square$

\section{Bulk universality for orthogonal and symplectic ensembles}\label{s:3}

In a remarkable paper \cite{Tr-Wi:98} Tracy and Widom
showed how to express the functions $S_{n,\beta}$
that appear in the definitions (\ref{Kn1}), (\ref{Kn4})
in terms of orthogonal polynomials defined by (\ref{w_n}) -- (\ref{orth}).
Set $\psi^{(n)}_j:=p_j^{(n)} \sqrt{w^{(n)}}$, $j\ge 0$. Then the system $\{\psi^{(n)}_j\}_{j\ge 0}$
defines an orthogonal basis in $L^2(\mathbb R)$ with respect to the standard inner product
$(f,g):= \int_{\mathbb R} f(\lambda)g(\lambda)\, d\lambda$. Moreover, they satisfy the recursion
relations
\begin{equation}\label{rec}
    \lambda\psi^{(n)}_{k}(\lambda)=a^{(n)}_{k+1}\psi^{(n)}_{k+1}(\lambda)+b^{(n)}_{k}\psi^{(n)}_{k}(\lambda)+
a^{(n)}_{k}\psi^{(n)}_{k-1}(\lambda),
\end{equation}
which define a semi-infinite Jacobi matrix $J^{(n)}$. It is known (see, e.g. \cite{PS:97}) that
\begin{equation}\label{b_J}
    |a^{(n)}_{k}|\le C,\quad |b^{(n)}_{k}|\le C,\quad |n-k|\le \varepsilon n.
\end{equation}

In order to state the formulae for $S_{n,\beta}$ we need to introduce more notation.
Let $D_\infty^{(n)}$ and $M_\infty^{(n)}$ be semi-infinite matrices
 that correspond to the differentiation
operator and to some integration operator respectively.
\begin{align*}
     D^{(n)}_\infty &:= \left(\left(\psi_j^{(n)}\right)',\psi_k^{(n)}\right)_{j,k\ge 0}\\
     M_\infty^{(n)} &:= \left (\epsilon \psi_j^{(n)},\psi_k^{(n)}\right)_{j,k\ge 0},\
     \textrm{with}\ (\epsilon f)(\lambda):=\int_{\mathbb R}\epsilon (\lambda-\mu)f(\mu)\, d\mu.
\end{align*}
Both matrices $D_\infty^{(n)}$ and $M_\infty^{(n)}$ are skew-symmetric.
Using in addition that for $j<k$
\begin{eqnarray*}
     (D_\infty^{(n)})_{jk}&=&\int_{\mathbb R} \left((p_j^{(n)}(\lambda) )'-
     \frac{n}{2} V'(\lambda)p_j^{(n)}(\lambda)\right)
     p_k^{(n)}(\lambda) w^{(n)}(\lambda) d\lambda\\&=&-\frac{n}{2}\int_{\mathbb R}V'(\lambda)p_j^{(n)}(\lambda)
     p_k^{(n)}(\lambda) w^{(n)}(\lambda) d\lambda
     =-\frac{n}{2}V'(J^{(n)})_{jk}\quad
\end{eqnarray*}
by orthogonality and the spectral theorem, we see  $(D_\infty^{(n)})_{j,k}=0$
for $|j-k|\ge 2m$ and
\[|(D_\infty^{(n)})_{j,k}|\le nC,\quad |j-n|, |k-n|\le \varepsilon n.\]
In particular, we may write
\begin{equation*}
     (\psi_j^{(n)})'= \sum_{|k-j|<2m} (D_\infty^{(n)})_{jk} \psi_k^{(n)}
\end{equation*}
as a finite sum. Since $\epsilon (\psi_j^{(n)})'=\psi_j^{(n)}$ we have for any $j,l\ge 0$ that
\begin{equation*}
     \delta_{jl}=(\epsilon (\psi_j^{(n)})',\psi_l)=\sum_{|k-j|<2m} (D_\infty^{(n)})_{jk}(M_\infty^{(n)})_{kl}.
\end{equation*}
This relation together with the skew-symmetry of $M_\infty^{(n)}$ and $D_\infty^{(n)}$ proves
\begin{equation*}
     D_\infty^{(n)} M_\infty^{(n)}=1 =M_\infty^{(n)} D_\infty^{(n)}.
\end{equation*}
Next we denote by $M_n^{(n)}$, $D_n^{(n)}$ the principal $n\times n$ submatrices of $M_\infty^{(n)}$ and $D_\infty^{(n)}$, i.e.
\begin{equation*}
     M_n^{(n)}:= ((M_\infty^{(n)} )_{jk})_{0\le j,k\le n-1}\quad,\quad D_n^{(n)}:= ((D_\infty^{(n)})_{jk})_{0\le j,k\le n-1}.
\end{equation*}
The formula of Tracy-Widom for $S_{n,\beta}$ now reads
\begin{align}\label{Sn1}
     S_{n,1}(\lambda,\mu) &= -\sum_{j,k=0}^{n-1} \psi_j^{(n)}(\lambda) (M_n^{(n)})_{jk}^{-1} (\epsilon \psi_k^{(n)})(\mu)\\
     \label{Sn4}
     S_{n/2,4}(\lambda,\mu) &= -\sum_{j,k=0}^{n-1} (\psi_j^{(n)})'(\lambda) (D_n^{(n)})_{jk}^{-1}  \psi_k^{(n)}(\mu)
\end{align}
As a by product of the calculation in \cite{Tr-Wi:98} one also obtains relations between the partition
functions $Q_{n,\beta}$ and the determinants of $M_n^{(n)}$ and $D_n^{(n)}$.
\begin{equation*}
     \det M_n^{(n)} =\left(\frac{Q_{n,1}\Gamma_n}{n!2^{n/2}}\right)^2\quad ,\quad
     \det D^{(n)}_n =\left(\frac{Q_{n/2,4}\Gamma_n}{(n/2)!2^{n/2}}\right)^2,
\end{equation*}
where $\Gamma_n:= \prod_{j=0}^{n-1} \gamma_j^{(n)}$ and $\gamma_j^{(n)}$ is the leading coefficient
of $p_j^{(n)}$. It is also known (see \cite{Me:91})
 that $Q_{n,2}=\Gamma_n^2/n!$ and we arrive at
\begin{equation*}
     \det (D_n^{(n)}M_n^{(n)}) =\left(\frac{Q_{n,1}Q_{n/2,4}}{Q_{n,2}(n/2)!2^n}\right)^2.
\end{equation*}
Since $ D_\infty^{(n)}M_\infty^{(n)}=1$ and $(D_\infty^{(n)})_{jk}=0$ for
$|j-k|>2m-1$ we have $D_n^{(n)}M_n^{(n)} =1 +\Delta_n$ with $\Delta_n$ being zero except
for the bottom $2m-1$ rows. Define $T_n$ to be the
$(2m-1)\times (2m-1)$ block in the bottom right corner of $D_n^{(n)}M_n^{(n)} $, i.e.
\begin{equation*}
     (T_n)_{jk}:= (D_n^{(n)}M_n^{(n)})_{n-2m+j, n-2m+k}\quad,\quad 1\le j, k\le 2m-1.
\end{equation*}
Then we have that $\det (T_n)$ equals $\det (M_n^{(n)}D_n^{(n)})$ and we arrive at a formula,
first observed by Stojanovic in \cite{St1}:
\begin{equation}\label{St}
     \det (T_n)= \left(\frac{Q_{n,1}Q_{n/2,4}}{Q_{n,2}(n/2)!2^n}\right)^2.
\end{equation}
Since $D_n^{(n)}M_n^{(n)}$ equals $1$ up to the matrix $\Delta_n$
of rank $2m-1$ (independent of $n$) it is conceivable that one may express $(M_n^{(n)})^{-1}$ and
$(D_n^{(n)})^{-1}$ that appear in (\ref{Sn1}), (\ref{Sn4}) by $D_n^{(n)}$
and $B_n^{(n)}$ respectively up to some correction terms that involves the inverse of $T_n^{-1}$. Using this idea Widom
provided in \cite{Wi:99} a useful formula for $S_{n,\beta}$ that was later refined in \cite{DGKV:07}.
In order to present this formula introduce some more notation:
\begin{align*}
     \Phi_1^{(n)} &:= (\psi_{n-2m+1}^{(n)} , \psi_{n-2m+2}^{(n)},\ldots ,\psi_{n-1}^{(n)})^T,\\
     \Phi_2^{(n)} &:= (\psi_n^{(n)},\psi_{n+1}^{(n)},\ldots , \psi_{n+2m-2}^{(n)})^T
\end{align*}
and
\begin{equation*}
     M_{rs} := (\epsilon \Phi_r^{(n)}, (\Phi_s^{(n)})^T),\qquad D_{rs}:=((\Phi_r^{(n)})',
     (\Phi_s^{(n)})^T),\quad 1\le r,s\le 2
\end{equation*}
define some $(2m-1)\times (2m-1)$ submatrices of $M_\infty^{(n)}$ and $D_\infty^{(n)}$.
Observe that $M_\infty^{(n)}D_\infty^{(n)}=1$ together with $(D_\infty^{(n)})_{jk}=0$
for $|j-k|\ge 2m$ implies
\begin{equation*}
     T_n=1-D_{12}M_{21}.
\end{equation*}
Finally we denote by $K_n(\lambda,\mu):= \sum^{n-1}_{j=0}\psi_j^{(n)} (\lambda) \psi_j^{(n)}(\mu)$
the reproducing kernel. We then have \cite{DGKV:07}
\begin{align*}
     S_{n,1}(\lambda,\mu) &= K_n(\lambda,\mu)+\Phi_1(\lambda)^TD_{12}\epsilon \Phi_2(\mu)-\Phi_1(\lambda)^T\hat G\epsilon \Phi_1(\mu),\\
     \hat G &:=D_{12}M_{22} (1-D_{21}M_{12})^{-1}D_{21}\\
     S_{n/2,4}(\lambda,\mu)&= K_n (\lambda,\mu) +\Phi_2(\lambda)^TD_{12} \epsilon \Phi_1(\mu)-\Phi_2(\lambda)^TG\epsilon \Phi_2(\mu),\\
     G&:= -D_{21}(1-M_{12}D_{21})^{-1}M_{11}D_{12}
\end{align*}
Since $S_{n,\beta}(\lambda,\mu)=-S_{n,\beta}(\mu,\lambda)$ one has for even $n$
\begin{equation*}
     (\epsilon S_{n,1})(\lambda,\mu)=-\int_\lambda^\mu S_{n,1}(t,\mu)\, dt ,
     \qquad(\epsilon S_{n/2,4})(\lambda,\mu)=-\int_\lambda^\mu S_{n/2,4}(t,\mu)\, dt.
\end{equation*}
Using this representation in the $21$-entry of $K_{n,\beta}$ together
with $\det T_n=\det (1-D_{21}M_{12})=\det (1-D_{12}M_{21})$ its straightforward to see that Theorem
\ref{Tt:2} follows from the following Lemma.

\begin{lemma}\label{Tl:2}
Given any compact set $K\subset (a,b)$ there exists a $C>0$ such that for all $n\ge 2m$ and all
$j,k\in \{n-2m+1,\ldots ,n+2m-2\}$ one has
\begin{align*}
(a)&  \quad\sup_{x\in K}|\epsilon \psi_j^{(n)}(\lambda)|\le
\frac{C}{\sqrt{n}};
&(b)& \quad |(M_\infty^{(n)})_{jk}|\le\frac{C}{n};
&(c)& \quad |\log \det (T_n)|\le C.
\end{align*}
\end{lemma}
Statements (a) and (b) will be derived from the asymptotics of the orthogonal polynomials in Appendix \ref{s:ap1}.
We now prove statement (c) using Theorem 1.

Consider  the functions $V_t$ defined in (\ref{V_t})
Then, as it was mentioned above (see the proof of Theorem \ref{t:log})
the limiting equilibrium density $\rho_t$ has the form  (\ref{rho_t}).
Hence, for any $t\in[0,1]$ $V_t$ satisfies conditions C1-C3 and if we
introduce the matrix $T_n(t)$ by the same way as above for the potential
$V_t$, then $T_n(0)$ corresponds to the GOE and GSE.
Consider the function
\begin{equation}\label{L}
    L(t)=\log\hbox{det}\,T_n(t).
\end{equation}
To prove that $|L(1)|\le C$ it is enough to prove that
\begin{equation}\label{cond_L}
|L(0)|\le C,\quad |L'(t)|\le C, \quad t\in[0,1]
\end{equation}
The first inequality here follows from the results of
\cite{Tr-Wi:98}. To prove the second inequality we use (\ref{St}) for $V$ replaced by $V_t$. Then we get
\[
L'(t)=n^2\int \Delta V(\lambda)p^{(n/2)}_{1,4,t}(\lambda)d\lambda
+n^2\int \Delta V(\lambda)p^{(n)}_{1,1,t}(\lambda)d\lambda
-2n^2\int \Delta V(\lambda)p^{(n)}_{1,2,t}(\lambda)d\lambda.
\]
It is easy to see that
\[
\lim_{n\to\infty}p^{(n/2)}_{1,4,t}=\lim_{n\to\infty}p^{(n)}_{1,1,t}=
\lim_{n\to\infty}p^{(n)}_{1,2,t}=\rho_t(\lambda)
\]
with $\rho_t$  defined by (\ref{rho_t}). Hence, using  (\ref{t1.1}), we obtain
 that the first and the second terms of (\ref{t1.1}) give zero contributions in
$L'(t)$, and therefore
\[
L'(t)=2(r_{n/2,4,t}(\Delta V)+r_{n,1,t}(\Delta V)-r_{n,2,t}(\Delta V)).
\]
But, according to Theorem \ref{t:1}, all terms here are bounded uniformly in $n$. Thus, we
have proved the second inequality in (\ref{cond_L}) and so statement (c) of Lemma \ref{Tl:2}.

$\square$

 \section{Appendix: uniform bounds for  $(M_\infty^{(n)})_{ij}$}\label{s:ap1}
 Set
 \begin{equation}\label{de_n}
    \delta_n=n^{-2/3+\kappa},\quad 0<\kappa<1/3.
\end{equation}
Then, according to  \cite{DKMVZ:99}, we have
\begin{eqnarray}\label{as_psi}
  \psi_n^{(n)}(\lambda)&=&\frac{\cos nF_n(\lambda)}{(4-\lambda^2)^{1/4}}\;(1+\mathcal{O}(n^{-1})),\quad|\lambda|\le 2-\delta_n;
  \notag\\
 \psi_{n-1}^{(n)}(\lambda)&=&\frac{\cos nF_{n-1}(\lambda)}{(4-\lambda^2)^{1/4}}\;(1+\mathcal{O}(n^{-1})),
 \quad|\lambda|\le 2-\delta_n;
 \notag\\
\psi_n^{(n)}(\lambda)&=&n^{1/6}B_{11}^{(\pm)}Ai\left(\pm n^{2/3}
\Phi_{\pm}(\lambda\mp 2)\right)(1+\mathcal{O}(|\lambda\mp2|))\\&&+n^{-1/6}B_{12}^{(\pm)}Ai'\left(\pm n^{2/3}
\Phi_{\pm}(\lambda\mp 2)\right)(1+\mathcal{O}(|\lambda\mp2|))+\mathcal{O}(n^{-1}),\quad |\lambda\mp2|\le\delta_n;\notag\\
\psi_{n-1}^{(n)}(\lambda)&=&n^{1/6}B_{21}^{(\pm)}Ai\left(\pm n^{2/3}
\Phi_{\pm}(\lambda\mp 2)\right)(1+\mathcal{O}(|\lambda\mp2|))\notag\\&&+n^{-1/6}B_{22}^{(\pm)}Ai'\left((\pm n^{2/3}
\Phi_{\pm}(\lambda\mp 2)\right)(1+\mathcal{O}(|\lambda\mp2|))+\mathcal{O}(n^{-1}),\quad |\lambda\mp2|\le\delta_n;\notag\\
|\psi_n^{(n)}(\lambda)|&\le& e^{-nc(|\lambda|- 2)^{3/2}},\quad
|\psi_{n-1}^{(n)}(\lambda)|\le e^{-nc(|\lambda|- 2)^{3/2}},\quad|\lambda|>2+\delta_n.\notag
\end{eqnarray}
where
\begin{equation}\label{F_n}
F_n(\lambda)=\frac{1}{2}\int_{\lambda}^2P(\lambda)\sqrt{4-\lambda^2}d\lambda+\frac{1}{n}\arccos(\lambda/2),\quad
F_{n-1}(\lambda)=F_n(\lambda)-\frac{2}{n}\arccos(\lambda/2)
\end{equation}
with $P$ defined in (\ref{P}).  Functions $\Phi_{\pm}$ in (\ref{as_psi}) are analytic in  some neighborhood of $0$
and  such that $\Phi_{\pm}(\lambda)= a_{\pm}x+\mathcal{O}(x^2)$ with some positive $a_{\pm}$.

Denote
\begin{eqnarray}\label{AB}
A(\lambda)&=&\frac{\sin nF_n(\lambda)}{n}\mathbf{1}_{|\lambda|\le2-\delta_n}+
\frac{\sin nF_n(2-\delta_n)}{n}\mathbf{1}_{|\lambda-2|\le\delta_n}+
\frac{\sin nF_n(-2+\delta_n)}{n}\mathbf{1}_{|\lambda+2|\le\delta_n}\notag\\
&&n^{-1/2}\left(\Psi\left(n^{2/3}\Phi_{+}(\lambda-2)\right)-\Psi\left(n^{2/3}\Phi_{+}(-\delta_n)\right)\right)
\mathbf{1}_{|\lambda-2|\le\delta_n}\notag\\
&&
-n^{-1/2}\left(\Psi\left(-n^{2/3}\Phi_{-}(\lambda+2)\right)
-\Psi\left(-n^{2/3}\Phi_{-}(\delta_n)\right)\right)
\mathbf{1}_{|\lambda+2|\le\delta_n}\\
B(\lambda)&=&\frac{1}{F_n'(\lambda)X^{1/4}(\lambda)}\mathbf{1}_{|\lambda|\le2-\delta_n}+
\frac{B_{11}^{(+)}}{\Phi_{+}'(\lambda-2)}\mathbf{1}_{|\lambda-2|\le\delta_n}+
\frac{B_{11}^{(-)}}{\Phi_{-}'(\lambda+2)}\mathbf{1}_{|\lambda+2|\le\delta_n}\notag
\end{eqnarray}
with
\begin{equation}\label{Psi}
\Psi(x):=\int_{-\infty}^x Ai(t)dt.
\end{equation}
\begin{proposition}\label{p:*} Under conditions of \textbf{C1-C3} for any smooth function $f$  we have
uniformly in $[-\delta_n-2,\delta_n+2]$
\begin{equation}\label{as_eps}
  \epsilon (f\psi_n^{(n)})(\lambda)=A(\lambda)B(\lambda)f(\lambda)+\epsilon r_n(\lambda)+\mathcal{O}(n^{-1})+
  \mathbf{1}_{|\lambda\pm 2|\le\delta_n}\mathcal{O}(n^{-5/6}),
\end{equation}
where
\[\int_{-2-\delta_n}^{2+\delta_n}|r_n(\lambda)|d\lambda\le Cn^{-1/2-3\kappa/4}.
\]
Similar representation
is valid for $\epsilon (f\psi_{n-1}^{(n)})$ if we replace in (\ref{AB}) $F_n$ by $F_{n-1}$
and $B_{11}^{(\pm)}$ by $B_{21}^{(\pm)}$.
Moreover, it follows from (\ref{as_eps}) that
\begin{equation}\label{b_eps}
 | \epsilon (f\psi_n^{(n)})(\lambda)|\le Cn^{-1/2},\quad| \epsilon (f\psi_{n-1}^{(n)})(\lambda)|\le Cn^{-1/2}.
\end{equation}
\end{proposition}
\textbf{Proof.}
We use the following simple relation, valid for any continuous piecewise differentiable functions
$A$, $f$, and any  piecewise differentiable $B$, if $A(\lambda)B(\lambda)f(\lambda)\to 0$, as $\lambda\to\pm\infty$:
\begin{equation*}\label{eps(AB)}
    \epsilon (A'Bf)(\lambda)=A(\lambda)B(\lambda)f(\lambda)-\epsilon (A(Bf)')(\lambda),
\end{equation*}
where $(Bf)'$ may contain $\delta$-functions at the points of jumps of $B$.
By the choice of $A,B$ (cf (\ref{AB}) and (\ref{as_psi})) $\epsilon (A'Bf)$ corresponds to the principal
part of $\epsilon (f\psi_n^{(n)})$. The terms $\mathcal{O}(n^{-1})$ and $\mathbf{1}_{|\lambda\pm 2|\le\delta_n}\mathcal{O}(n^{-5/6})$
  in (\ref{as_eps}) appear because of the integrals of $\mathcal{O}(n^{-1})$ in the second line of (\ref{as_psi}) and
  the terms in the forth line of (\ref{as_psi}) respectively. Hence we need only to prove the bound for
  $r_n=A(\lambda)(Bf)'$.  Observe that
\begin{eqnarray*}\int|r_n(\lambda)|d\lambda &\le&
\frac{C}{n^{1/2}}\left(\int_{2-\delta_n}^{2+\delta_n}+\int_{-2-\delta_n}^{-2+\delta_n}\right)|(Bf)'(\lambda)|d\lambda
\\&&+\frac{1}{n}\int_{-2+\delta_n}^{2-\delta_n}|
(Bf)'(\lambda)|d\lambda=n^{-1/2}\mathcal{O}(\delta_n)+n^{-1}\mathcal{O}(\delta_n^{-3/4})=\mathcal{O}(n^{-1/2-3\kappa/4}).
\end{eqnarray*}
$\square$

Using recursion relations (\ref{rec})
it is easy to get that for any $|j|\le 2m$
\[\psi^{(n)}_{n+j}(\lambda)=f_{0j}(\lambda)\psi^{(n)}_{n}(\lambda)+f_{1j}(\lambda)\psi^{(n)}_{n-1}(\lambda),\]
where $f_{0j}$ and $f_{1j}$ are polynomials of degree at most $|j|$. Note that
since it is known that $a^{(n)}_{k}$ and $b^{(n)}_{k}$ for $k-n=o(n)$ are bounded uniformly in $n$,
$f_{0j}$ and $f_{1j}$ have  coefficients, bounded uniformly in $n$. Hence
for our purposes  it is enough to estimate
\begin{equation}\label{int}
I_1:=(f_{1j}\psi^{(n)}_{n-1},\epsilon(f_{0k}\psi^{(n)}_{n})),\;
I_2:=(f_{1j}\psi^{(n)}_{n-1},\epsilon(f_{1k}\psi^{(n)}_{n-1})),\;
I_3:=(f_{0j}\psi^{(n)}_{n},\epsilon(f_{0k}\psi^{(n)}_{n})).
\end{equation}
  It follows from Proposition \ref{p:*} that
  \[I_1=I_{11}+I_{12}+I_{13}+
(f_{1j}\psi^{(n)}_{n-1},\,\epsilon r_n)+\mathcal{O}(n^{-1}),\]
where
\begin{eqnarray*}
I_{11}&=&n^{-1}\int_{-2+\delta_n}^{2-\delta_n}\frac{f_{0j}(\lambda)f_{1k}(\lambda)\sin nF_n(\lambda)
\cos nF_{n-1}(\lambda)}
{F'_n(\lambda)(4-\lambda^2)^{1/2}}d\lambda,\\
I_{12}&=&
n^{-1/3}B^{(+)}_{11}B^{(+)}_{21}\int_{2-\delta_n}^{2+\delta_n}
\left(\Psi\left(n^{2/3}\Phi_{+}(\lambda-2)\right)-\Psi\left(n^{2/3}\Phi_{+}(-\delta_n)\right)\right)\\
&&\cdot\frac{Ai\left(n^{2/3}\Phi_{+}(\lambda-2)\right)}{\Phi_{+}'(\lambda-2)}f_{0j}(\lambda)f_{1k}(\lambda)d\lambda,
\end{eqnarray*}
and $I_{13}$ is the integral similar to $I_{12}$ for the region $|\lambda+2|\le\delta_n$. It is easy to see
that
\begin{eqnarray*}
I_{12}&=&B^{(+)}_{11}B^{(+)}_{21}\frac{f_{0j}(2)f_{1k}(2)}{2n(\Phi_{+}'(0))^2}\,\left(1+o(1)\right),\\
I_{13}&=&B^{(-)}_{11}B^{(-)}_{21}\frac{f_{0j}(-2)f_{1k}(-2)}{2n(\Phi_{+}'(0))^2}\,\left(1+o(1)\right)
\end{eqnarray*}
Moreover, using the bound for $r_n$ from (\ref{as_eps}) and (\ref{b_eps}), we get
\[(f_{1j}\psi^{(n)}_{n-1},\epsilon r_n)=-(\epsilon(f_{1j}\psi^{(n)}_{n-1}), r_n)\le
\int|\epsilon(f_{1j}\psi^{(n)}_{n-1})|\,|r_n|\,d\lambda=\mathcal{O}(n^{-1-3\kappa/4}).\]
Hence we are left to find the bound for $I_{11}$.
\begin{eqnarray*}
I_{11}&=&(2n)^{-1}\int_{-2+\delta_n}^{2-\delta_n}\frac{f_{0j}(\lambda)f_{1k}(\lambda)\sin n(F_n(\lambda)-F_{n-1}(\lambda))}
{F'_n(\lambda)(4-\lambda^2)^{1/2}}d\lambda\\&&
+(2n)^{-1}\int_{-2+\delta_n}^{2-\delta_n}\frac{f_{0j}(\lambda)f_{1k}(\lambda)\sin n(F_n(\lambda)+F_{n-1}(\lambda))}
{F'_n(\lambda)(4-\lambda^2)^{1/2}}d\lambda=I_{11}'+I_{11}''
\end{eqnarray*}
By the definition of $F_n$ and $F_{n-1}$ (\ref{F_n}), we obtain
\[I_{11}'=\frac{1+o(1)}{2n}\int_{-2}^{2}\frac{f_{0j}(\lambda)f_{1k}(\lambda)}
{P(\lambda)(4-\lambda^2)^{1/2}}d\lambda.\]
Moreover, integrating by parts one can get easily that $I_{11}''=\mathcal{O}(n^{-2}\delta_n^{-3/2})=\mathcal{O}(n^{-1-3\kappa/2})$.

The other two integrals from (\ref{int}) can be estimated similarly.

$\square$

 \section{Appendix: proof of the bounds (\ref{b1_de})}\label{s:ap2}
Let us introduce a function $H$ which we call Hamiltonian to stress the analogy with statistical mechanics.
\begin{equation*}\label{H}
    H(\Lambda)=-\sum_{i=1}^{n}(V(\lambda_i)+n^{-1}h(\lambda_i))+2\sum_{1\le i<j\le n}\log|\lambda_i-\lambda_i|,
    \quad \Lambda=(\lambda_1,\dots\lambda_n).
\end{equation*}
It is evident that for any continuous $f(\lambda_1,\dots,\lambda_k)$
\[\int f(\lambda_1,\dots,\lambda_k)p^{(n)}_{k,\beta,h}(\lambda_1,\dots,\lambda_k)d\lambda_1\dots d\lambda_k=
\frac{\int f(\lambda_1,\dots,\lambda_k)e^{n\beta H(\Lambda)}d\Lambda}
{\int e^{n\beta H(\Lambda)}d\Lambda}=:\langle f\rangle_{\beta H}
\]
Moreover we introduce the "approximating" Hamiltonian, depending on a functional
parameter $m:\hbox{supp }m\subset [-2,2]$
\begin{equation*}  \label{IDS.2.8}
H_{a}(\Lambda ;m)=\sum_{i=1}^{n}v_n(\lambda_i;m)+(n-1)\mathcal{L}[m,m].
\end{equation*}%
Here
\begin{eqnarray}  \label{v_n}
v_{n}(\lambda ;m)&=&-V(\lambda )-\frac{1}{n}h(\lambda )+2\frac{n-1}{n}\mathcal{L}(\lambda ;m),\\
\mathcal{L}(\lambda ;m)&=&\int\log|\lambda-\mu|m(\mu)d\mu,\notag\\
\mathcal{L}[m,m]&=&\int\log|\lambda-\mu|^{-1}
m(\lambda)m(\mu)d\lambda d\mu.
\notag
\end{eqnarray}%
By the Jensen inequality for any two real functions $\mathcal{H}_{1}(\Lambda)$,
$\mathcal{H}_{2}(\Lambda)$ we have
\begin{eqnarray*}
\frac{\int e^{n\beta\mathcal{H}_{1}(\Lambda)/2}d\Lambda}{\int
e^{n\beta\mathcal{H}_{2}(\Lambda)/2}d\Lambda}\ge e^{n\beta/2\left\langle
\mathcal{H}_{1}-\mathcal{H}_{2}\right\rangle_{\beta\mathcal{H}_{2}}},\quad
\frac{\int e^{n\beta\mathcal{H}_{2}(\Lambda)/2}d\Lambda}{\int
e^{n\beta\mathcal{H}_{1}(\Lambda)/2}d\Lambda}\ge e^{n\beta/2\left\langle
\mathcal{H}_{2}-\mathcal{H}_{1}\right\rangle_{\beta\mathcal{H}_{2}}}
\end{eqnarray*}
where we denote $\langle\dots\rangle_{\beta\mathcal{H}_{\delta}}=\int(\dots)e^{n\beta\mathcal{H}_{\delta}
(\Lambda)/2}d\Lambda/
\int e^{n\beta\mathcal{H}_{\delta}(\Lambda)/2}d\Lambda$ ($\delta=1,2$).
Then we get
\begin{equation*}\label{Bog}
\left\langle
\mathcal{H}_{2}-\mathcal{H}_{1}\right\rangle_{\beta\mathcal{H}_{1}}\le
\left\langle\mathcal{H}_{2}-\mathcal{H}_{1}\right\rangle_{\beta\mathcal{H}_{2}}
\end{equation*}
Taking here $\mathcal{H}_{1}=H$, $\mathcal{H}_{2}=H_{a}$, we obtain
\begin{equation}\label{Bog.1}
R[m]:= \frac{\int (H_{a}-H)
e^{-\beta nH/2}d\Lambda }{(n-1)\int e^{-\beta nH/2}d\Lambda}\le
\frac{\int (H_{a}-H)e^{-\beta n H_{a}(\Lambda ;m)/2}d\Lambda  }{(n-1)\int e^{-\beta n H_{a}(\Lambda ;m)/2}d\Lambda}
=:R_{a}[m] ,%
\end{equation}%
 Since $H$ and $H_{a}$ are
symmetric, we can rewrite the l.h.s. of (\ref{Bog.1}) as
\begin{multline}
R[m]=\int \log \frac{1}{|\lambda -\mu |}
\bigg(p_{2,\beta,h }^{(n)}(\lambda ,\mu )-p_{1,\beta,h }^{(n)}(\lambda )p_{1,\beta,h }^{(n)}(\mu
)\bigg)d\lambda d\mu +\mathcal{L}[p_{1,\beta,h }^{(n)}-m,p_{1,\beta,h }^{(n)}-m],
\label{IDS.R}
\end{multline}%
where $p_{1,\beta,h }^{(n)}$ and $p_{2,\beta,h }^{(n)}$ are defined
by (\ref{p_nl}) if we replace $V$ by $V_h$. To obtain the expression for the r.h.s. of (\ref{Bog.1}) we need to
replace $p_{2,\beta }^{(n)}(\lambda )$ and $p_{2,\beta }^{(n)}(\lambda ,\mu )$ in (\ref{IDS.R}) by
$p_{1,\beta,h}^{(n,a)}(\lambda ;m)$ and $p_{1,\beta,h}^{(n,a)}(\lambda ;m)p_{1,\beta,h}^{(n,a)}(\mu ;m)$,
-- correlation functions of the ~approximating
Hamiltonian (\ref{IDS.2.8}), where
\begin{equation}  \label{rho_a}
p_{1,\beta,h}^{(n,a)}(\lambda ;m)=e^{\beta nv_n(\lambda;m )/2} \bigg(\int
d\lambda e^{\beta nv_n(\lambda;m )/2}\bigg)^{-1}.
\end{equation}%
This yields:
\begin{equation}
R_{a}[m]=\mathcal{L}[p_{1,\beta,h}^{(n,a)}-m,p_{1,\beta,h}^{(n,a)}-m],  \label{Ram}
\end{equation}%
Now let us choose the function $m$. Set
\begin{equation}\label{m_n}
   m_n(\lambda)=\frac{n}{n-1}\left(\rho(\lambda)+\frac{1}{\beta n}\nu_n(\lambda)\right)\mathbf{1}_{|\lambda|\le 2},
\end{equation}
where
\begin{equation}\label{nu_n}
 \nu_n(\lambda)= \frac{\sqrt{4-\lambda^2}}{\pi}\int_{-2}^2d\mu\frac{
((\log\rho)_n'(\mu)+\beta h'(\mu)/2)}
{(\mu-\lambda)\sqrt{4-\mu^2}}+\frac{\alpha_n}{\pi\sqrt{4-\lambda^2}}=\nu_n^{(1)}(\lambda)+\alpha_n\nu^{(0)}(\lambda),
\end{equation}
 the function $(\log\rho)_n(\lambda)$   coincides with $\log\rho(\lambda)$ on the interval
$\sigma_n=[-2+n^{-1/2},2-n^{-1/2}]$,
and  $(\log\rho)_n(\lambda)$ is a linear function for $\lambda\in\sigma\setminus\sigma_n$, chosen so that
$(\log\rho)_n(\lambda)$ has continuous derivative on $\sigma$. The constant $\alpha_n$ here  is chosen to provide the
condition
\begin{eqnarray*}\int m_n(\lambda)d\lambda=1\Longleftrightarrow \alpha_n&=&-\beta-\int\nu_n^{(1)}(\lambda)d\lambda
\\&=&
 -\beta-\int_{-2}^2\frac{((\log\rho)_n'(\mu)+\beta h'(\mu)/2)\mu d\mu}{\sqrt{4-\lambda^2}}=\mathcal{O}(n^{1/4}).
 \end{eqnarray*}
Since $\rho$ has the form (\ref{rho}),  $\nu_n^{(1)}$ is a sum of a bounded function
which comes from $P$ and of a negative function which comes from  the integral of $(\log\sqrt{4-\lambda^2})'$. Hence,
\begin{equation}\label{|nu|}
    \int|\nu_n^{(1)}(\lambda)|d\lambda\le C-\int\nu_n^{(1)}(\lambda)d\lambda=\mathcal{O}(n^{1/4}).
\end{equation}
It is easy to see that $\nu_n(\lambda)$ is chosen to satisfy the equation
\[
  \int_{-2}^2\frac{\nu_n(\mu)d\mu}{\lambda-\mu}=(\log\rho)_n'(\lambda)+\beta h'(\lambda)/2.
\]
Therefore
\begin{equation}\label{A.1}
  \mathcal{L}(\lambda,\nu_n)=(\log\rho)_n(\lambda)+\beta h(\lambda)/2+r_n(\lambda),
\end{equation}
where  for $|\lambda|\le 2$ $r_n(\lambda)=C_n$ and $C_n$ is a constant independent of $\lambda$, but depending on $n$.
One can find $C_n$ as
\begin{equation}\label{A.3}
  C_n=\mathcal{L}(0,\nu_n)-(\log\rho)_n(0)-\beta h(0)/2=
  \int\log|\lambda|\nu_n(\lambda)d\lambda-(\log\rho)_n(0)-\beta h(0)/2=\mathcal{O}(n^{1/4})
\end{equation}
(here we used  that $\nu_n^{(1)}$ for $|\lambda|\le1$ is bounded uniformly in $n$ ).
Hence,
\begin{equation}\label{A.2}
 \beta n v_n(\lambda,m_n)/2=\beta n\bigg(\mathcal{L}(\lambda,\rho)-V(\lambda)\bigg)/2+(\log\rho)_n(\lambda)+C_n,
  \quad |\lambda|\le 2
\end{equation}
Let us estimate $\frac{d}{d\lambda}\mathcal{L}(\lambda,\nu_n^{(1)})$ for $\lambda>2$.
From (\ref{nu_n}) we get
\begin{multline*}\label{b_der}
%\frac{\nu_n(\lambda)}{\sqrt{4-\lambda^2}}
\left|\frac{d}{d\lambda}\mathcal{L}(\lambda,\nu_n^{(1)})\right|=\bigg|
\int\frac{\nu_n^{(1)}(\lambda_1)d\lambda_1}{\lambda-\lambda_1}
\bigg|=\bigg|\frac{1}{\pi}\int_{-2}^2d\lambda_1\int_{-2}^2d\mu\frac{\sqrt{4-\lambda^2_1}
((\log\rho)_n'(\mu)+\beta h'(\mu)/2)}
{(\lambda-\lambda_1)(\mu-\lambda_1)\sqrt{4-\mu^2}}\bigg|\\
=\bigg|\int_{-2}^2\bigg(1-\frac{\sqrt{\lambda^2-4}}{\lambda-\mu}\bigg)\frac{((\log\rho)_n'(\mu)+\beta h'(\mu)/2)}
{\sqrt{4-\mu^2}}d\mu\bigg|\\
\le\sup\{|(\log\rho)_n'(\mu)+\beta h'(\mu)/2|\}\int_{-2}^2d\mu\bigg(1+\frac{\sqrt{\lambda^2-4}}{\lambda-\mu}\bigg)
(4-\mu^2)^{-1/2}d\mu\le n^{1/4}C_1.
\end{multline*}
Here we  used the identities (valid for $\lambda\not\in[-2,2]$)
\[\frac{1}{\pi}\int_{-2}^2d\lambda_1\frac{\sqrt{4-\lambda^2_1}}
{(\lambda-\lambda_1)(\mu-\lambda_1)}=1-\frac{\sqrt{\lambda^2-4}}{\lambda-\mu},\quad
\frac{1}{\pi}\int_{-2}^2\frac{(4-\mu^2)^{-1/2}d\mu}{\lambda-\mu}=\frac{1}{\sqrt{\lambda^2-4}},
\]
and the bound $\sup|(\log\rho)_n'(\mu)|\le Cn^{1/4}$. Therefore
\[\mathcal{L}(\lambda,\nu_n)-\mathcal{L}(2,\nu_n)\le C_1n^{1/4}|\lambda-2|+ C_2n^{1/4}|\lambda-2|^{1/2},
\]
where the second term in the l.h.s. comes from $\alpha_n\nu^{(0)}$ in (\ref{m_n}).
Moreover, since under conditions C2, C3  there exists $C^*$
such that
\[\mathcal{L}(\lambda,\rho)-V(\lambda)=C^*,\, |\lambda|\le 2,\quad
\mathcal{L}(\lambda,\rho)-V(\lambda)-C^*\le-C_0|\lambda^2-4|^{3/2},\, |\lambda|\ge 2,
\]
 we obtain
\begin{eqnarray*}
n\beta v_n(\lambda,m_n)/2-n\beta C^*/2-C_n=(\log\rho)_n\mathbf{1}_{|\lambda|\le 2}
+\tilde r_n(\lambda)\mathbf{1}_{|\lambda|> 2},\\
\tilde r_n(\lambda)\le Cn^{1/4}(|\lambda|-2)^{1/2}-nC_0|\lambda^2-4|^{3/2}.
\end{eqnarray*}
Then we have
\begin{equation}\label{int_v}
\int_{\mathbb{R}\setminus\sigma} e^{n\beta v_n(\lambda,m_n)/2-n\beta C^*/2-C_n}d\lambda\le
2\int_0^{\infty}e^{Cn^{1/4}x^{1/2}-C_0nx^{3/2}} dx\le Cn^{-2/3}.
\end{equation}
The last bound  can be obtained by splitting the interval $[0,\infty)$ in two parts:
$[0,n^{-2/3})$ and $[n^{-2/3},\infty)$. Then in the first interval we used the fact
that $\sup\{Cn^{1/4}x^{1/2}-nx^{3/2}\}\le c$ and in the second interval
we used that this function is negative, its derivative is a negative decreasing function, bounded from above by
$(-Cn^{2/3})$.

For $|\lambda|\le 2$, since $|e^{(\log \rho)_n}-\rho|\le Cn^{-1/4}\mathbf{1}_{\sigma\setminus\sigma_n}$ we get
\[
\int_{-2}^2e^{(\log\rho)_n}d\lambda=\int_{\sigma_n}\rho(\lambda)d\lambda+\mathcal{O}(n^{-3/4})=1+\mathcal{O}(n^{-3/4}).
\]
Hence, using the above inequality and (\ref{int_v}), we obtain
\begin{equation*}\label{A.4a}
p_{1,\beta,h}^{(n,a)}(\lambda ;m_n)=\frac{\rho(\lambda)+\mathcal{O}(n^{-1/4})\mathbf{1}_{\sigma\setminus\sigma_n}}{1+\mathcal{O}(n^{-2/3})}=
\rho(\lambda)+\mathcal{O}(n^{-1/4})\mathbf{1}_{\sigma\setminus\sigma_n}+\mathcal{O}(n^{-2/3}).
\end{equation*}
Thus
\begin{equation*}\label{A.4}
p_{1,\beta,h}^{(n,a)}(\lambda ;m_n)-m_n=-\frac{\beta}{n}\nu_n(\lambda)+
\mathcal{O}(n^{-1/4})\mathbf{1}_{\sigma\setminus\sigma_n},
+\mathcal{O}(n^{-2/3})
\end{equation*}
and
\[
\mathcal{L}[p_{1,\beta,h}^{(n,a)}(\lambda ;m_n)-m_n,p_{1,\beta,h}^{(n,a)}(\lambda ;m_n)-m_n]\le
C\left(n^{-2}\mathcal{L}[\nu_n,\nu_n]+n^{-4/3}+n^{-3/2}\log n\right).
\]
Moreover, using (\ref{A.1}) and (\ref{|nu|}), we write
\begin{equation}\label{A.5}
\mathcal{L}[\nu_n,\nu_n]=-\int\bigg((\log\rho)_n(\lambda)+\beta h(\lambda)/2+C_n\bigg)\nu_n(\lambda)
d\lambda\le Cn^{1/2}.
\end{equation}
Finally we get
\begin{equation*}\label{A.6}
\mathcal{L}[p_{1,\beta,h}^{(n,a)}(\lambda ;m_n)-m_n,p_{1,\beta,h}^{(n,a)}(\lambda ;m_n)-m_n]\le
Cn^{-4/3}.
\end{equation*}
The inequality combined with (\ref{Bog.1}) gives us
\begin{eqnarray}\label{A.7}
&&\hskip-2cm\int \log \frac{1}{|\lambda -\mu |}
\bigg(p_{2,\beta,h }^{(n)}(\lambda ,\mu )-p_{1,\beta,h }^{(n)}(\lambda )p_{1,\beta,h }^{(n)}(\mu
)\bigg)d\lambda d\mu \\&&\hskip2cm +\mathcal{L}[p_{1,\beta,h }^{(n)}-m_n,p_{1,\beta,h }^{(n)}-m_n]\le
Cn^{-4/3}.
\notag\end{eqnarray}
Let us prove that
\begin{equation}\label{A.8}
\int \log \frac{1}{|\lambda -\mu |}
\bigg(p_{2,\beta,h }^{(n)}(\lambda ,\mu )-p_{1,\beta,h }^{(n)}(\lambda )p_{1,\beta,h }^{(n)}(\mu
)\bigg)d\lambda d\mu\ge -C\log n/n.
\end{equation}
Introduce the function
\[l_n(\lambda)=\log|\lambda|^{-1}\mathbf{1}_{|\lambda|>n^{-7}}+
\left(\log n^7+n^{7}(n^{-7}-|\lambda|)\right)\mathbf{1}_{|\lambda|<n^{-7}}.
\]
It is easy to check that for any $k\not=0$ the Fourier transform $\widehat l(k)\ge 0$
and hence, for any positive operator $K:L_2(\mathbb{R})\to L_2(\mathbb{R})$  and such that $\int
K(\lambda,\mu)d\lambda d\mu=0$, we have
\begin{equation}\label{A.9}
\int l_n(\lambda -\mu)K(\lambda,\mu)d\lambda d\mu\ge 0
\end{equation}

From (\ref{p_nl}) it is easy to obtain that for any $|x|\le n^{-3}$
\[|p_{1,\beta,h }^{(n)}(\lambda +x )-p_{1,\beta,h }^{(n)}(\lambda  )|\le
Cn^{-1}(\sup|V'|+\sup|h'|),\]
therefore
\[1\ge\int p_{1,\beta,h }^{(n)}(\lambda)d\lambda\ge(1-Cn^{-1}(\sup|V'|+n^{-1}\sup|h'|))\max p_{1,\beta,h }^{(n)}(\lambda)
n^{-3}.\]
Hence
\[\max p_{1,\beta,h }^{(n)}(\lambda)\le n^3(1+Cn^{-1}(\sup|V'|+n^{-1}\sup|h'|)).\]
Similarly
\[\max p_{2,\beta,h }^{(n)}(\lambda,\mu)\le n^6(1+Cn^{-1}(\sup|V'|+n^{-1}\sup|h'|))\]
Then,  the above bounds and the inequality
\[\int|\log \frac{1}{|\lambda -\mu |}-l_n(\lambda -\mu)|d\lambda\le Cn^{-7}\log n,\]
imply
\begin{eqnarray*}
&&\hskip-0.5cm\int \log \frac{1}{|\lambda -\mu |}
\bigg(p_{2,\beta,h }^{(n)}(\lambda ,\mu )-p_{1,\beta,h }^{(n)}(\lambda )p_{1,\beta,h }^{(n)}(\mu
)\bigg)d\lambda d\mu \hskip 7cm\\ &&\ge\int l_n(\lambda -\mu)
\bigg(p_{2,\beta,h }^{(n)}(\lambda ,\mu )-p_{1,\beta,h }^{(n)}(\lambda )p_{1,\beta,h }^{(n)}(\mu
)\bigg)d\lambda d\mu-\mathcal{O}(n^{-1}\log n)\\
&&\ge\int l_n(\lambda -\mu)
\bigg(p_{2,\beta,h }^{(n)}(\lambda ,\mu )-\frac{n}{n-1}p_{1,\beta,h }^{(n)}(\lambda )p_{1,\beta,h }^{(n)}(\mu
)\bigg)d\lambda d\mu-\mathcal{O}(n^{-1}\log n)\hskip 1cm\\
&&=\frac{1}{n(n-1)}\int l_n(\lambda -\mu)k_n(\lambda,\mu)d\lambda d\mu
-\frac{1}{n-1}\int l_n(0)p_{1,\beta,h }^{(n)}(\lambda )d\lambda-\mathcal{O}(n^{-1}\log n),
\end{eqnarray*}
where the kernel $k_n$ is defined by (\ref{k_n}). Since $k_n$ is positively definite, and $\int
k(\lambda,\mu)d\lambda d\mu=0$, we can use (\ref{A.9}), and taking into account that
$l_n(0)=\mathcal{O}(\log n)$, obtain (\ref{A.8}).

Then
\[
\bigg(\mathcal{L}^{1/2}[p_{1,\beta,h }^{(n)}-\rho,p_{1,\beta,h
}^{(n)}-\rho]-\mathcal{L}^{1/2}[m_n-\rho,m_n-\rho]\bigg)^2
\le\mathcal{L}[p_{1,\beta,h }^{(n)}-m_n,p_{1,\beta,h
}^{(n)}-m_n]\le C\log n/n
\]
And since it follows from (\ref{m_n}) and (\ref{A.5}) that
\[\mathcal{L}[m_n-\rho,m_n-\rho]\le Cn^{-4/3},\]
we have
\begin{equation}\label{A.10}
\mathcal{L}[p_{1,\beta,h }^{(n)}-\rho,p_{1,\beta,h
}^{(n)}-\rho]\le Cn^{-1}\log n.
\end{equation}
For any $\Im z\not=y$, taking the Fourier transforms $\widehat p_{1,\beta,h }^{(n)}$ and
$\widehat\rho$ of the functions $p_{1,\beta,h }^{(n)}$ and $\rho$, we get
\begin{multline*}
\bigg|\int\frac{p_{1,\beta,h }^{(n)}(\lambda)-\rho(\lambda)}{\lambda-z}d\lambda\bigg|
\le2\int_0^\infty|\widehat p_{1,\beta,h }^{(n)}(k)-\widehat\rho(k)| e^{-|k||y|}dk\\
\le 2\bigg(\int |k| e^{-|k||y|}dk\bigg)^{1/2}\bigg(\int\frac{|\widehat p_{1,\beta,h
}^{(n)}(k)-\widehat\rho(k)|^2}{k}dk\bigg)^{1/2}=2|y|^{-1}\mathcal{L}^{1/2}[p_{1,\beta,h }^{(n)}-
\rho,p_{1,\beta,h}^{(n)}-\rho].
\end{multline*}
Then   (\ref{A.10}) yields  the second and the third bounds of (\ref{b1_de}). The first bound
follows from Lemma \ref{l:1}.

\end{document}